\documentclass[twocolumn,showpacs,preprintnumbers,amsmath,amssymb,prl]{revtex4}

\usepackage{graphicx}
\usepackage{dcolumn}
\usepackage{bm}

\begin{document}

\preprint{}

\title{Internal state of granular assemblies near random close packing}

\author{Jean-No\"el Roux}
\email{Jean-Noel.Roux@lcpc.fr}
\affiliation{%
Laboratoire des Mat\'eriaux et des Structures du G\'enie Civil\\
2 all\'ee Kepler, Cit\'e Descartes, 77420 Champs-sur-Marne, France
}%
\date{\today}
\begin{abstract}
The structure of random sphere packings in mechanical equilibrium in prescribed
stress states, as studied by molecular dynamics simulations, strongly depends on
the assembling procedure. Frictionless packings in the limit of low pressure are devoid
of dilatancy, and consequently share the same
random close packing density, but exhibit fabric anisotropy related to stress
anisotropy. Efficient compaction methods can be viewed as routes 
to circumvent the influence of
friction. Simulations designed to resemble two such procedures, lubrication and vibration (or ``tapping'') show that 
the resulting granular structures differ, the less dense one having, remarkably, the larger coordination
number. Density, coordination number and fabric can thus
vary independently. Calculations of elastic moduli and comparisons with experimental results suggest
that measurable elastic properties provide information on those important internal state
variables.
\end{abstract}

\pacs{45.70.-n, 45.70.Cc, 71.55.Jv, 81.40.Jj}
\keywords{granular packings, disordered materials, elasticity}
\maketitle
The mechanical properties of solidlike granular packings and their microscopic, grain-level origins
are an active field of research in material science and condensed-matter physics~\cite{HHL98,KI01},
with practical motivations in soil mechanics and material processing, as well as
theoretical ones, as general approaches to the rheology of ``jammed'' systems~\cite{LN01} are attempted.
For long, the only accessible information on the internal state 
of granular packings has been the
density or the solid volume fraction $\Phi$. Its importance was recognized, \emph{e.g.} in soil mechanics~\cite{DMWood,HHL98},
and for the elaboration of concrete mixes~\cite{FdL99}. The concept of
\emph{random close packing} (RCP)~\cite{CC87} is deemed
relevant in many contexts. For monodisperse spheres, the RCP volume
fraction value $\Phi_{RCP} \simeq 0.637$ was consistently reported~\cite{CC87,BH93}. 
Discrete numerical simulation~\cite{CUND79} proved a valuable tool to investigate the internal state of packings, as it is
capable to reproduce mechanical behaviors~\cite{TH00}, 
and to identify relevant variables other than $\Phi$, such as coordination number and fabric
(or distribution of contact orientations)~\cite{BR90,RR01}. 
The influence of the sample preparation procedure on the mechanics of
solid granulates is widely recognized as crucial, in numerical experiments as
well as in real ones, but it is seldom discussed in detail. Experimentally, a variety of techniques
such as controlled rain deposition under gravity~\cite{RT87}, or layerwise tamping are used. Numerically, loose contactless
configurations (``granular gases'') are homogeneously compressed~\cite{TH00,MGJS99,MJS00}
or dropped in a container under their weight~\cite{SEGHL02,SGL02}. In both cases,
configurations are usually classified by their density, other state parameters
being implicitly regarded as density-dependent. 
The present Letter reports on a molecular dynamics study that further investigates those initial states and their dependence on the
assembling procedure in the case of monodisperse spherical particles.
Its aim is twofold: first to revisit (after others~\cite{KTS02}) the prevailing notion of
``random close packing''; then to study the internal states obtained by procedures designed to approach such 
RCP states, and to suggest possible means to measure important hidden information on their geometry.

A foreword is necessary about the
\emph{mechanical} definition~\cite{JNR2000} of a configuration of maximum density. 
Let us consider a collection of a large number of identical rigid spheres, and enclose them in a rectangular parallelipipedic
box.
The degrees of freedom are the positions 
of sphere centers and the lengths $L^\alpha$ ($\alpha=1,\, 2, \, 3$) of the container edges parallel to coordinate
axes (changes of cell shape might also be considered). Using arbitrary reference lengths $L_0^{\alpha}$ 
strain parameters $\epsilon _{\alpha} = \frac{L_0^{\alpha} -L^{\alpha} }{L_0^{\alpha}}$ are defined.
A local minimum of volume $V$, or maximum of $\Phi$, is obtained on minimizing, under impenetrability constraints,
the potential energy $W=-P\sum_\alpha \epsilon_{\alpha}$ 
that corresponds to external pressure $P>0$. Constraints entail the definition of repulsive, normal contact forces as
Lagrange parameters. Local maxima of $\Phi$ in configuration space (called ``strictly jammed states'' in~\cite{KTS02})
are thus \emph{equivalently} characterized as stable equilibrium configurations of rigid, frictionless particles
under an isotropic pressure.
This duality between contact forces and rigid constraints, presented \emph{e.g.,} in~\cite{JNR2000},
leads to \emph{define} any such configuration, devoid of crystal nucleus (which might be checked
with suitable order parameters~\cite{VCKB02}), as a random close-packed state. 
Not surprisingly, it is common practice, in order
to obtain dense states easily, to set friction coefficients to zero in  molecular dynamics
calculations~\cite{MGJS99,TH00}. 

Real materials are made of elastic, deformable grains, a feature most often taken into account in
simulations~\cite{CUND79,TH00,MGJS99}.
In those presented below, we implement the Hertz
law~\cite{JO85} for normal forces, and, in the presence of friction, like in refs.~\cite{MGJS99,SEGHL02,SGL02},
a simplified form of the Cattaneo-Mindlin-Deresiewicz relations~\cite{JO85}
(as suggested in~\cite{EB96}) for tangential forces.
Stable equilibria
then minimize a potential energy containing an additional elastic term. Solutions for
perfectly rigid grains are recovered~\cite{JNR2000} in the limit of large contact stiffness, relative to the level
of externally imposed forces. Specifically, the typical ratio $h/a$ of elastic normal deflection of
surfaces in contact $h$ to sphere diameter $a$ is of order $(P/E)^{2/3}$ for Hertzian contacts
(force $F \sim E a^{1/2} h^{3/2}$). 
Our simulations, to allow for comparisons with experiments, correspond
to glass beads with Young modulus $E=70kPa$ and Poisson coefficient $\nu=0.3$. 
The method is similar to that of refs.~\cite{CUND79,TH00,MGJS99}
except that stresses, rather than strain rates, are controlled (as in ref.~\cite{PR82}). 
Unless otherwise specified,
results are averaged over 5 samples of 4000 beads enclosed in a periodic cubic cell.

Compression of a frictionless granular gas to equilibrium under $P=10kPa$ yields
$\Phi= 0.637\pm 0.002 \simeq \Phi_{RCP}$, like in refs.~\cite{MGJS99,OSLN03}.
The difference with the value
$\Phi^*$ one would obtain in the $P\to 0$ limit might be estimated to fisrt order
in the small elastic deflections $h_c$ of the contacts, via the theorem of virtual work~\cite{JNR2000},
on equating the work of the external pressure, $PV (\Phi-\Phi^*)/ \Phi$ to the work of normal
contact forces $f_c$, $\sum_c f_c h_c$, between equilibrium configurations at $P=0$ and $P=10kPa$.
One finds $\Phi-\Phi^* \simeq 1.15\times 10^{-4}$, less than statistical uncertainties on $\Phi$.
Configurations in the $P\to 0$ limit are endowed with 
particular properties related to absence of force indeterminacy and stability,
which together entail the isostaticity of the force-carrying structure~\cite{JNR2000}, which consequently 
has a coordination number  $z^*=6$. On the force network at $10kPa$ each ball has an average of
$z^*=6.073\pm 0.004$ contacts, in agreement with refs.~\cite{MGJS99,OSLN03,SEGHL02}.
This value excludes ``rattlers'', spheres that carry no force,
which represent a fraction $f_0=(1.3\pm 0.3)\%$ of the total number, and differs from 
the global coordination number, $z=z^*(1-f_0)$ (obtained on attributing zero contact to rattlers). 
In all cases considered, only a few ( $0.3\%$) isolated  particles belong to crystalline regions,
as defined in ref.~\cite{VCKB02}, p. 4.

\begin{figure}
\includegraphics[height=6cm]{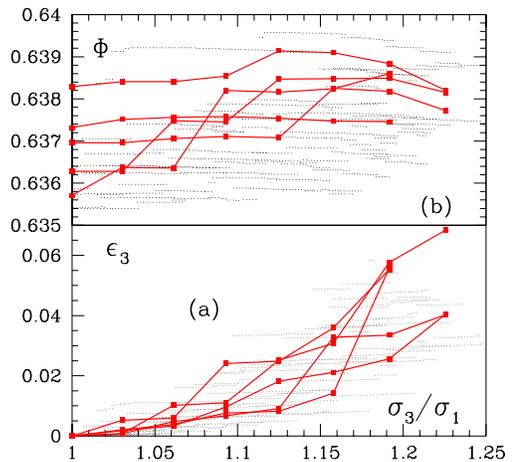}
\caption{\label{fig:anisf1} (Color online) (a) Axial strain $\epsilon_3$ and (b) volume fraction $\Phi$ vs.
stress ratio $\sigma_3/\sigma_1$. Small symbols: earlier simulations, 1372 balls. Connected dots:
4000 ball samples.}
\end{figure}

\begin{figure}
\includegraphics[height=6cm]{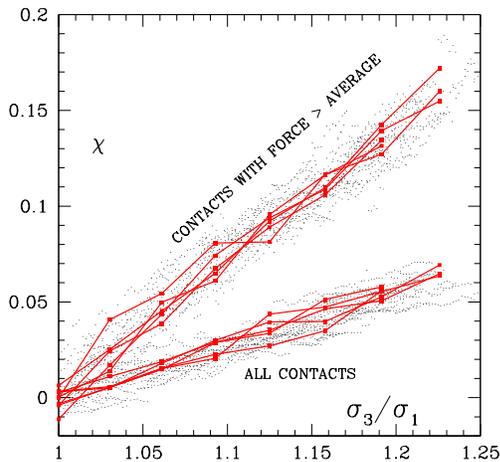}
\caption{\label{fig:anisf2} (Color online). Fabric parameter $\chi$ versus principal stress ratio. Symbols as on
fig.~\ref{fig:anisf1}. $\chi$ is larger for contacts carrying larger than average forces. Note the apparent regression of
fluctuations as sample size increases, unlike on fig.~\ref{fig:anisf1}.}
\end{figure}

Numerical configurations made with friction are known to exhibit lower densities
and coordination numbers~\cite{TH00,MJS00,OSLN03,SGL02}.
In practice, {\em compaction strategies should therefore avoid the mobilization of friction in contacts}. 
Our definition of RCP states is naturally associated with an isotropic pressure. What about other possible states of stress ?
To investigate those, isotropic samples of frictionless were submitted to stepwise
\emph{axisymmetric triaxial compression }, \emph{i.e.,} 
keeping the coordinate axes as principal stress directions, $\sigma_3$ was increased,
at constant $P=(\sigma_1+\sigma_2+\sigma_3)/3$,
by steps of $0.02P$, while maintaining 
$\sigma_2=\sigma_1$. At each stress step 
one waits for mechanical equilibrium (to accuracy $10^{-4}a^2P$ on the net force on each particle). With perfectly
rigid balls, one would have a local minimum of $W=-\sum_{\alpha} \sigma_{\alpha}\epsilon _{\alpha}$. 
The maximum principal stress ratio supported by frictionless balls was observed to reach $1.2$ to $1.24$. For each equilibrium,
$\Phi$, $z^*$, $f_0$, strains $\epsilon _{\alpha}$ (taking the initial isotropic state as reference),
and the fabric parameter $\chi = <3\cos^2\theta-1>$ ($\theta$ is the angle of normals to contacts with the major principal stress 
direction $z$) are recorded. Figure~\ref{fig:anisf1} displays $\epsilon _3$ (the \emph{axial} strain)
and $\Phi$ as functions of $\sigma_3/\sigma_1$. 
Stress-strain relations are elusive in those frictionless systems, as axial strain responds quite erratically
to deviator stress increments (the strain/stress curve was
interpreted as the trajectory of a L\'evy flight in 2D~\cite{CR2000}).
$\Phi$ variations with deviator stress are extremely small, with a slight increasing trend, so that observed densities
remain within the small interval of reported RCP values. This lack of dilatancy, contrary to the classical
Reynods~\cite{RE85} argument, might appear as a paradox, as
density should first
increase on destabilizing density-minimizing configurations. Such density increases 
are however likely to vanish in the large system limit.
This lack of volume change agrees with experimental observations, as stresses in alleged
RCP states (\emph{e.g.}, under gravity) cannot be always isotropic. Likewise, the numerical simulations that should 
produce anisotropic stress states, due to a final oedometric compression~\cite{MJS00}, or to gravity~\cite{SEGHL02,SGL02}, 
nevertheless yield $\Phi\simeq \Phi_{RCP}$ in the frictionless case.
On the other hand, one observes (fig.~\ref{fig:anisf2}) a gradual building of a moderate
contact network anisotropy, apparently determined by stress anisotropy for macroscopic samples (a likely consequence of the
``fragility'' of equilibrium states~\cite{CR2000}).
The ratio of vertical (parallel to axis $3$, $C_{33}$) to horizontal ($C_{11}$) 
computed oedometric elastic moduli reach about $1.2$ at the largest deviator.
Consequently, assuming full success of the compaction method
in avoiding friction mobilization, observed RCP configurations belong to a continuum of different states. Those
differ at least by one internal variable: fabric anisotropy, related
to stress anisotropy, which should be accessible via measurements of elastic constants.

Let us now turn to systems in which the effects of friction are not entirely suppressed, and investigate, focussing on
isotropic states, the
possible effects of two different compacting strategies. One is lubrication, 
either with a viscous, fluid oil (hydrodynamic lubrication), or with a
very thin, greasy solid layer coating the grain surfaces.  
The suppression of friction is then expected to be more efficient in the course of the assembling process
 than in the static packing under higher
loads. To produce numerical configurations bearing some similarity to experimental, lubricated ones, we simulated
slow, isotropic compressions, with a high level of viscous dissipation,
using a small friction coefficient $\mu_0=0.02$ up to equilibrium 
states under $P=10kPa$. To study the effect of higher pressures (maintaining isotropy), the friction coefficient was 
raised to $\mu=0.3$, a typical value for a dry contact. Such numerical configurations are referred to as B samples in
the sequel, A samples denoting those that were assembled without friction (note, however, that elastic moduli were subsequently
evaluated in the presence of tangential forces, assuming type A configurations,
well annealed at each pressure level, mobilize tangential elasticity~\cite{MGJS99} in response to small load increments).

Another common procedure to circumvent friction and produce dense samples consists in 
vibrating, shaking or ``tapping''~\cite{NKBJN98}. In practice, this is most often how experimental
RCP states are reached with dry beads, for which a certain amount of intergranular friction
is unavoidable. This might be understood, as the replacement of
permanent contacts by repeated collisions precludes the mobilization of friction, and therefore destabilizes arrangements
that were equilibrated thanks to tangential contact forces. A third series of numerical configurations (series C)
was produced in order to mimic the internal state of vibrated samples, as follows.
First, A samples at $P=10kPa$ were dilated, scaling all coordinates by a factor $1.005$, so that
all contacts opened. Then particles were attributed random velocities and
left to collide without energy dissipation (like the hard sphere fluid of ref.~\cite{VCKB02}) at constant
volume. This shaking stage lasted until each particle had undergone an average of 50 collisions. Samples were
finally isotropically compressed with $\mu=0.3$ and strongly dissipative dynamics, to equilibrium at $P=10kPa$.

\begin{figure}
\includegraphics[height=11cm]{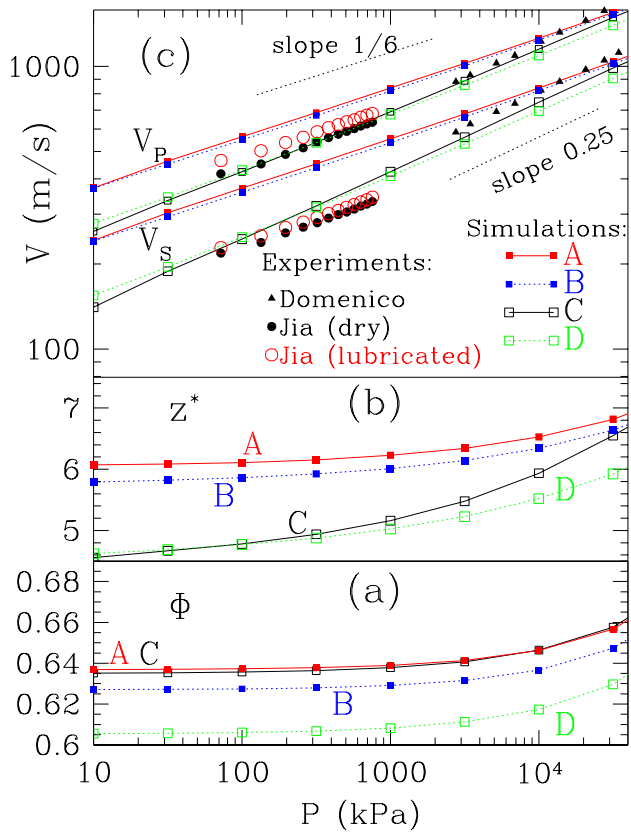}
\caption{\label{fig:phizvp} (Color online) (a) Solid fraction $\Phi$; (b) coordination number $z^*$; 
(c) speeds of sound $V_P$ and $V_S$ (with experimental points~\cite{DO77,JM01}) for samples of types A, B, C, D, versus 
isotropic pressure $P$. $V_P$ and $V_S$ correlate to $z^*$ rather than to $\Phi$. Error bars are smaller than symbols.}
\end{figure}

A comparison of low-pressure equilibrated configurations A, B and C reveals that the ``imperfectly lubricated'' samples B
do not quite reach the RCP solid fraction, as $\Phi=0.627\pm 0.0002$ is observed at $10kPa$. However,
the coordination number
of active grains, $z^*=5.79\pm 0.007$, was only slightly reduced and the proportion of rattlers, $(1.7\pm 0.2)\%$, rather similar.
On the other hand, ``vibrated'' samples C are denser, with $\Phi=0.635\pm 0.001$ very close to $\Phi_{RCP}$, but their coordination
number is considerably smaller, $z^*=4.56\pm 0.03$, with $f_0=(13.3\pm 0.6)\%$.
Remarkably, this shows that density and coordination number
can vary independently for the same material, without fabric anisotropy. In fact, $z^*$ values in the C case are as low as
in loose samples (series D, $z^*=4.62\pm 0.01$, $\Phi=0.606\pm 0.002$ at $10kPa$) 
prepared from a cold ``granular gas'' by direct isotropic compression with $\mu=0.3$ (see fig.~\ref{fig:phizvp}).
Compressions to higher pressures (keeping $\mu=0.3$) 
and calculations of elastic moduli enabled comparisons to experimental results. On figure~\ref{fig:phizvp},
$z^*$, $f_0$, $\Phi$ and velocities of longitudinal and transverse sound 
velocities, respectively denoted as $V_P$ and $V_S$, are plotted
versus $P$ in the $10kPa$ to $100MPa$ range. $V_P$ and $V_S$ are deduced from computed bulk moduli ($B$)
and shear moduli ($G$) as
$V_P=\sqrt{(B+\frac{4}{3}G)/\rho}$ and $V_S=\sqrt{G/\rho}$ ($\rho$ is the mass density of the packing).
Sound speeds are compared to experimental measurements on glass beads by Domenico~\cite{DO77},
and by Jia and Mills~\cite{JM01}. Numerical results with C (vibrated)
samples are clearly in better agreement with experimental results
on dense packings of dry beads, than A samples(\emph{i.e.}, initially
prepared in an ideal isotropic RCP state), for which $V_P$ and $V_S$ are too large in the $100kPa$ range, 
and vary too slowly with $P$. (Results for A samples are in excellent agreement with ref.~\cite{MGJS99}).
Sound speed dependence on $P$ can
be approximated with a power law, the exponent being roughly $1/6$ for A samples, as predicted by simple ``effective
medium'' approaches~\cite{MGJS99}, whereas it is
close to $0.25$ for $V_S$ and about $0.22$ for $V_P$ in C samples (due to faster variations of modulus $G$). Despite the
remaining uncertainties and discrepancies in the comparison to experiments (the numerical procedure is a caricature of
the experimental assembling process, and no information is available on sample anisotropies in~\cite{DO77,JM01}, while numerical
configurations are isotropic) we conclude that the numerical simulations of dense samples of dry particles should
probably involve such annealing stages, which result in much smaller coordination numbers. 

Interestingly, Jia \emph{et al.}~\cite{JM01} also prepared packings with a very small amount of a solid lubricant (trioleine)
and observed 
\emph{smaller} densities than for dry packings (typically $\Phi\simeq 0.625$ instead of $0.64$ at $100Pa$),
but \emph{larger} sound speeds (fig.~\ref{fig:phizvp}). 
Our results for B samples (``imperfect lubrication'') are qualitatively similar, as we find
sound speeds close to A (ideal RCP) values, larger than C (``vibrated'') ones.
Note also the agreement for the slope of $\log(V)$ versus $\log(P)$ (close to 1/6).
The larger 
elastic moduli of initially lubricated samples might thus be attributed to their higher coordination number
(in spite of their lower density).

To conclude, we propose to define \emph{ideal} random close packings of spherical balls
as equilibrium states devoid of crystal nuclei that remain stable
without friction, and to regard compaction procedures as recipes to suppress friction mobilization. Simulations reveal
that, even though they all share the same volume fraction and coordination
number (in the small stress limit),
such RCP states differ by at least another state variable, a fabric tensor: 
they develop anisotropic contact networks to support anisotropic stress states.
Simple modelling of two \emph{imperfect} compacting strategies, vibration and lubrication,
show that isotropic configurations with very different coordination numbers coexist
close to the RCP density. Solid fraction, coordination number and fabric can vary independently.
Comparisons with laboratory measurements
of sound speeds under varying pressure on glass bead samples show good correspondence with real assembling procedures,
and suggest that elastic properties provide access to coordination and fabric.

\end{document}